\documentclass[%
reprint,
amsmath,amssymb,
aps,
braket,
]{revtex4-2}

\usepackage{graphicx}
\usepackage{dcolumn}
\usepackage{braket}
\usepackage{bm}
\usepackage{color}

\begin{document}


\title{Doppler-free three-photon spectroscopy on narrow-line optical transitions}

\author{Guglielmo Panelli}%
\thanks{These authors contributed equally to this work.}%
\affiliation{%
 Department of Physics, Stanford University, Stanford CA, 94305 
}%
\author{Shaun C. Burd}
\thanks{These authors contributed equally to this work.}%
\affiliation{%
 Department of Physics, Stanford University, Stanford CA, 94305 
}%
\author{Erik J. Porter}%
\affiliation{%
 Department of Physics, Stanford University, Stanford CA, 94305 
}%

\author{Mark Kasevich}
\email{kasevich@stanford.edu}
\affiliation{%
 Department of Physics, Stanford University, Stanford CA, 94305 
}%

\date{\today}

\begin{abstract}
We demonstrate coherent Doppler-free three-photon excitation of the $^{1}S_{0}$\,$\leftrightarrow$\,$^{3}P_{0}$ optical clock transition and the $^{1}S_{0}$\,$\leftrightarrow$\,$^{3}P_{1}$ intercombination transition in free-space thermal clouds of $^{88}$Sr atoms. By appropriate orientation of the wavevectors of three lasers incident on the atoms, the first-order Doppler shift can be eliminated for all velocity classes. Three-photon excitation of the $^{1}S_{0}$\,$\leftrightarrow$\,$^{3}P_{1}$ transition enables high-contrast Ramsey spectroscopy with interrogation times comparable to the 21\,$\mu$s natural lifetime using a single near-resonant laser source. Three-photon spectroscopy on the $^{1}S_{0}$\,$\leftrightarrow$\,$^{3}P_{0}$ clock transition, using only laser frequencies nearly resonant with the $^{1}S_{0}$\,$\leftrightarrow$\,$^{3}P_{0}$ and $^{1}S_{0}$\,$\leftrightarrow$\,$^{3}P_{1}$ transitions, enables a reduction in Doppler broadening by two orders of magnitude and a corresponding $\sim470\,$Hz linewidth without a confining potential.

\end{abstract}

\maketitle


\section{Introduction}
Laser spectroscopy is an essential tool for understanding the nature of matter and for performing the most precise physical measurements. However, spectroscopic resolution can be limited by Doppler effects stemming from the motion of atoms and molecules. Furthermore, effects of momentum recoil from photon absorption pose limitations for many quantum sensing applications. Here we demonstrate a Doppler-free and recoil-free spectroscopic technique on narrow-line atomic transitions with wide applicability in spectroscopy and precision metrology.

Various techniques have been developed to mitigate Doppler effects and achieve high-resolution spectroscopy. One approach is to confine atomic motion within a spatial region much smaller than the wavelength of the probing radiation. This confinement virtually eliminates Doppler effects and is used in the most accurate optical atomic clocks: laser-cooled neutral atoms in optical lattices \cite{Bothwell_2019,Mcgrew2018} and ions in electrodynamic traps \cite{Brewer2019}. However, this confinement typically limits the number of atoms that can contribute to the spectroscopic signal, a factor in the stability of an atomic clock. 
On the other hand, atom-beam-based approaches using optical Ramsey spectroscopy \cite{ramsey1950molecular, Riehle1991} can employ a large atomic flux but are limited by residual Doppler and photon recoil effects \cite{olson2019ramsey}. Another technique that does not require particle confinement is saturated-absorption spectroscopy where a narrow velocity class within a thermal distribution is selectively targeted \cite{Lamb1964, McFarlane1963}. However this limits the number of atoms that can contribute to the signal. 
Two-photon spectroscopy enables participation of all velocity classes, but poses stringent requirements on the electronic structure of the atom. Moreover, this technique cannot be used for many important optical transitions because only states of the same parity can be coupled \cite{Goppert_Mayer1931,Abella1962} unless high optical intensities are employed \cite{Leanhardt2014}.

Three-photon excitation addresses these limitations, enabling Doppler-free excitation of odd-parity transitions with participation of all atoms in a thermal ensemble   \cite{cagnac1973spectroscopie,grynberg1980multiphoton}. This was first demonstrated by Grynberg et al. who performed Doppler-free three-photon spectroscopy on a dipole-allowed transition in a room temperature sodium vapor \cite{Grynberg1976}.  Doppler-free three-photon excitation has been shown to be a promising method for Rydberg excitation \cite{thoumany2009spectroscopy, Ryabtsev2011,Tretyakov2022, beterov2023three}, lasing without inversion \cite{Rein2021}, and for recoil-free quantum logic gates~\cite{zhang2024}.  Multispectral three-photon schemes have also been studied theoretically for narrow-line optical transitions \cite{Hong2005,Barker2016}, particularly in the context of optical atomic clocks \cite{Ludlow2015}. However, these proposals pose stringent atomic-species-dependent requirements along with challenging phase and frequency stabilization requirements for the three optical excitation lasers.

Here we report Doppler- and recoil-free three-photon excitation of the narrow-line $^{1}S_{0}$\,$\leftrightarrow$\,$^{3}P_{1}$ transition ($\Gamma/2\pi=7.6\,$  kHz natural linewidth) as well as the $^{1}S_{0}$\,$\leftrightarrow$\,$^{3}P_{0}$  forbidden clock transition in atomic strontium. 
In the first case, three near-resonant lasers derived from the same source are strategically oriented to cancel the Doppler effect. This enables temperature-insensitive Ramsey spectroscopy with coherence decay dominated by the lifetime of the transition. For Doppler-free spectroscopy on the clock transition, the third laser addresses the $^{3}P_{0}$~clock state enabling free-space Ramsey spectroscopy and a demonstrated $\sim$100--fold reduction in the spectroscopic linewidth for atoms probed in free space. Both schemes are widely generalizable to different atomic species and optical transitions.  

\begin{figure*}
\includegraphics[width = 6.5in]{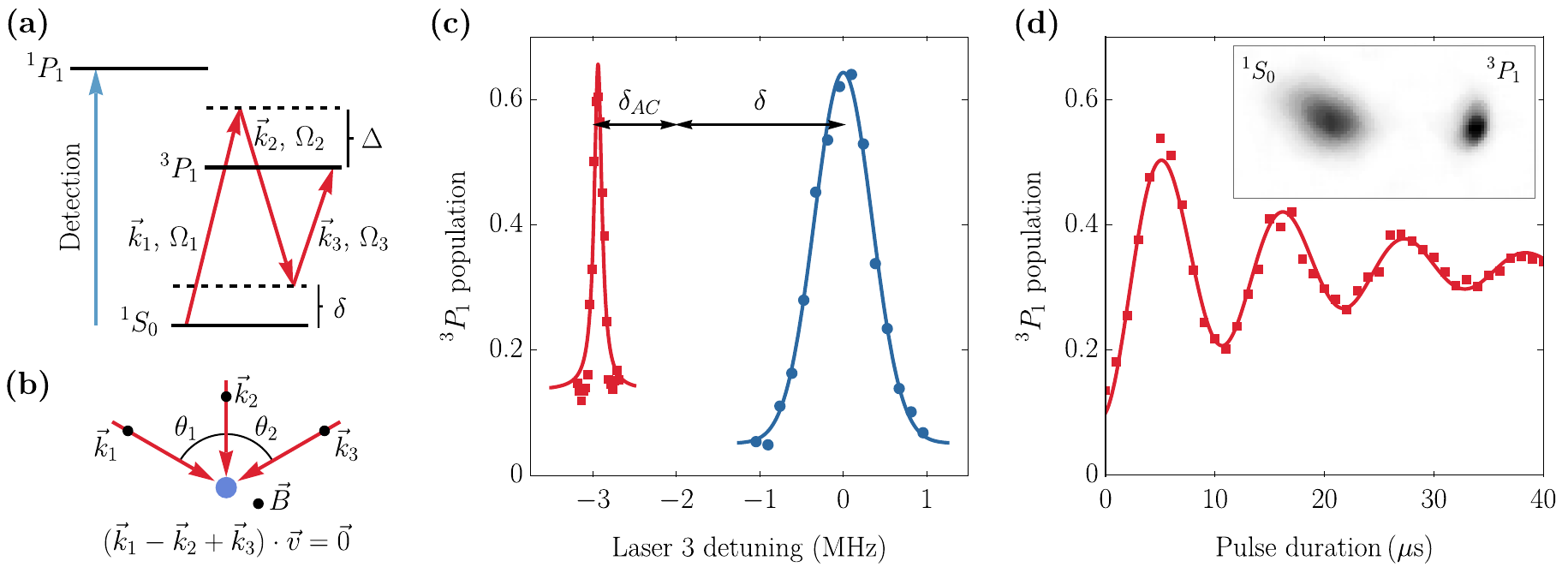}
\caption{\label{fig:fig689} Three-photon Doppler-free excitation of the $^{1}S_{0}$ $\leftrightarrow$\,$^{3}P_{1}$ transition in $^{88}$Sr. (a) Energy level diagram (not to scale) for three-photon excitation of $^{1}S_{0}$\,$\leftrightarrow$\,$^{3}P_{1}$. Red arrows indicate lasers coupling atomic states with the Rabi frequencies $\Omega_{i}$, $i\in \{1,2,3\}$, $\delta$ and $\Delta$ are frequency detunings of the lasers described above. (b) Laser geometry for Doppler-free operation. The three lasers have wavevectors $\vec{k}_{i}$, $i\in \{1,2,3\}$ and propagate in a plane, with lasers 1 and 2 separated by angle $\theta_{1}$ and lasers 2 and 3 separated by $\theta_{2}$. The laser polarization vectors (indicated by black dots) are perpendicular to the plane containing the laser propagation vectors and parallel to the applied magnetic field $\vec{B}$. (c) Three-photon (red squares) and single-photon (blue circles) resonance features, plotted as the population in $^{3}P_{1}$ versus the frequency of laser 3 relative to the single-photon $^{1}S_{0}$\,$\leftrightarrow$\,$^{3}P_{1}$ resonance (0\,MHz). Solid curves are Lorentzian (red) and Gaussian (blue) fits to the data. (d) Three-photon Rabi oscillations in the population of $^{3}P_{1}$ as the three-photon pulse duration is varied. The inset shows an example fluorescence image used for state readout. }
\end{figure*}

\section{Excitation of the ${^3}P_{1}$ Transition}
As illustrated in Fig.~\ref{fig:fig689}, three coplanar laser beams with wavevectors $\vec{k}_{1}, \vec{k}_{2}$, and $\vec{k}_{3} $ interact with an atomic cloud with Rabi frequencies $\Omega_{1}$, $\Omega_{2}$, and $\Omega_{3}$, respectively. In the Doppler-free configuration, the angles between the laser beams $\theta_{1}$ and $\theta_{2}$ are selected so that $\vec{k}_{1}-\vec{k}_{2}+\vec{k}_{3}=\vec{0}$. This ensures that the first-order Doppler shift $(\vec{k}_{1}-\vec{k}_{2}+\vec{k}_{3})\cdot \vec{v}~=~\vec{0}$, regardless of the velocity $\vec{v}$ of an atom. 
The angular frequencies of lasers 1 and 2 are given by $\omega_{1}=\omega_{0}+\Delta$ and $\omega_{2}=\omega_{0}+\Delta-\delta$ respectively, where $\omega_{0}$ is the resonance frequency of the $^{1}S_{0}$\,$\leftrightarrow$\,$^{3}P_{1}$ transition. For driving $^{1}S_{0}$\,$\leftrightarrow$\,$^{3}P_{1}$, the frequency of laser~3 is set to $\omega_{3}=\omega_{0}-\delta+\delta_{AC}$ where $\delta_{AC}$ accounts for the AC Stark shift of the transition due to the three-photon excitation lasers. For the $^{1}S_{0}$\,$\leftrightarrow$\,$^{3}P_{0}$ scheme, the frequency of laser 3 is $\omega_{3}=\omega_{\text{clock}}-\delta+\delta_{AC}$, where $\omega_{\text{clock}}$ is the resonance frequency of the $^{1}S_{0}$\,$\leftrightarrow$\,$^{3}P_{0}$ transition.
When tuned to the three-photon resonance frequency, with the conditions that $\Delta \gg \Omega_{1}$, $\Delta-\delta \gg \Omega_2$, and $\delta \gg \Omega_{3}$, the three-photon transition occurs with Rabi frequency $\Omega_{3\nu}$ given by  $\Omega_{3\nu} = \Omega_1 \Omega_2 \Omega_3/2\Delta \delta$ for the $^{1}S_{0}$\,$\leftrightarrow$\,$^{3}P_{1}$ transition and $\Omega_{3\nu} = \Omega_1 \Omega_2 \Omega_3/4\Delta \delta$ for the $^{1}S_{0}$\,$\leftrightarrow$\,$^{3}P_{0}$ transition~\cite{Ryabtsev2011}. The factor of two between these expressions arises from the two excitation paths available in the $^{1}S_{0}$\,$\leftrightarrow$\,$^{3}P_{1}$ excitation scheme, see Appendix A for details. 

We prepare clouds of $\sim$10$^6$ $^{88}$Sr atoms using a dual-stage magneto-optical trap (MOT) sequence, first with cooling on the 461\,nm $^{1}S_{0}$\,$\leftrightarrow$\,$^{1}P_{1}$ transition \cite{ludlow2008strontium} then on the narrow-line 689\,nm $^{1}S_{0}$\,$\leftrightarrow$\,$^{3}P_{1}$ transition \cite{Katori1999}, resulting in $\sim$3\,$\mu$K atomic ensembles. The 689\,nm light for the three-photon drive is generated from an injection-locked laser diode (Ushio HL69001DG). Seed light for the lock is sourced from a 689\,nm external-cavity diode laser locked to a high-finesse optical cavity (Stable Laser Systems). The light is split into three paths each with an independent acousto-optical modulator (AOM) for varying the laser frequency and intensity. The AOMs are controlled using a field-programmable gate array (FPGA) based timing system \cite{Bourdeauducq2016}. After the AOMs, each laser beam is coupled into a separate polarization-maintaining optical fiber. The outputs of the optical fibers are collimated to give a 0.8\,mm waist at the location of the atoms, which is $\sim\,$7 times the typical root-mean-square (rms) radius of the atom cloud of 120\,$\mu$m. For $^{1}S_{0}$\,$\leftrightarrow$\,$^{3}P_{1}$ three-photon excitation, the polarization vector of each laser is parallel to a $4$\,G quantization magnetic field (Fig.~\ref{fig:fig689}b) that is pulsed on during the interaction time in order to resonantly address only the $m_{J}=0$ Zeeman sublevel within the $^{3}P_{1}$ manifold.  At the end of a given experiment, a  1-2$\,\mu$s pulse resonant with the $^{1}S_{0}\leftrightarrow\, ^{1}P_{1}$ (461\,nm) transition followed by a 4\,ms delay is used to spatially separate atoms in states $^{1}S_{0}$ and $^{3}P_{1}$ after the three-photon interrogation (Fig.~\ref{fig:fig689}d,~inset). All atoms are in the ground state after the separation time delay and fluorescence imaging on the 461\,nm transition provides population readout. For readout of $^{3}P_{0}$ populations, an additional repumping pulse is applied during the cloud separation delay to bring all atoms in $^{3}P_{0}$ to $^{1}S_{0}$ before fluorescence detection. 

\begin{figure*}
\includegraphics[width=6.5in]{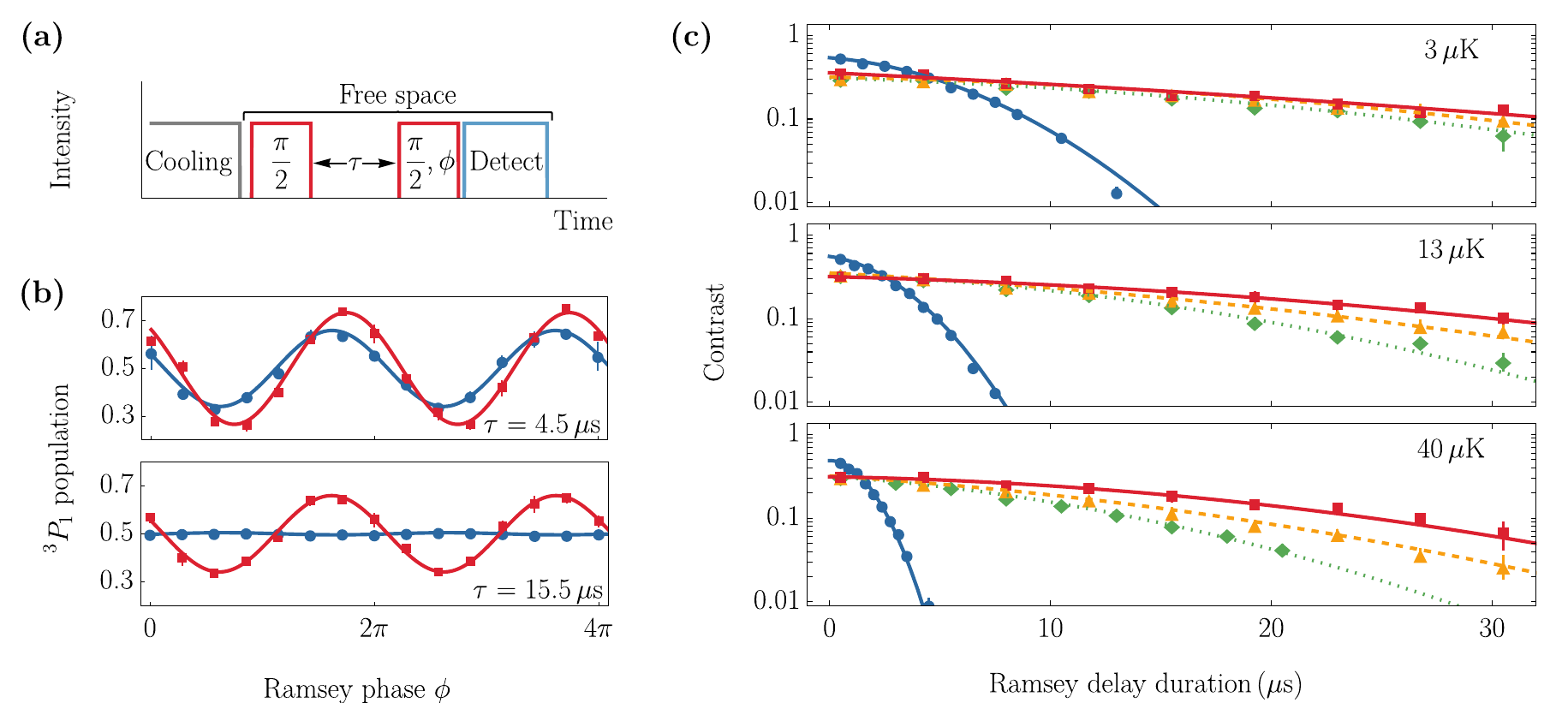}
\caption{\label{fig:fig689ramsey} Comparison of $^{1}S_{0}$\,$\leftrightarrow$\,$^{3}P_{1}$ single and three-photon Ramsey coherence. a) Experimental sequence. After preparation of the atoms, three-photon Ramsey $\pi/2$-pulses each with duration $2.5\,\mu$s and relative phase $\phi$ are applied to the atoms. The Ramsey pulses are separated by variable delay $\tau$. For single-photon excitation, only a single resonant laser with Rabi frequency $\Omega_{3}$ is used, with a $\pi/2$-pulse duration of $0.5\,\mu$s. b) Plots of the measured population of $^{3}P_{1}$ against $\phi$  with $\tau= 4.5\,\mu$s (upper panel) and  $\tau$ = 15.5\,$\mu$s (lower panel). Data points are for three-photon excitation with optimized laser beam geometry (red squares) and for single-photon  excitation (blue circles). Data points and error bars indicate the average and the standard error of 3 repeated measurements, respectively. The solid curves are sinusoidal fits to the data used to determine the Ramsey contrast. c) Ramsey contrast as $\tau$ is varied for single- and three-photon Ramsey experiments for atom ensemble temperatures of 3, 13, and 40\,$\mu$K. Three-photon Ramsey contrast data is given for experimentally optimized laser alignment (red squares), $\Delta\theta_{2}=4^{\circ}$ (yellow triangles), and $\Delta\theta_{1}=2.5^{\circ}$, $\Delta\theta_{2}=4^{\circ}$ (green diamonds). Blue circles indicate contrast measured using single-photon excitation. Error bars indicate the standard error of the contrast measurement. Lines are  fits to the data, see Appendix B for details. }
\end{figure*}

A typical $^{1}S_{0}$\,$\leftrightarrow$\,$^{3}P_{1}$ three-photon spectrum is shown in Fig.~\ref{fig:fig689}c. Single-photon Rabi frequencies of $\Omega_1 = 2.12(2)\,$MHz, $\Omega_2 = 3.12(3)\,$MHz, and $\Omega_3 = 0.54(4)\,$MHz, and detunings of $\Delta = 10.1\,$MHz and $\delta = 2.0\,$MHz were used, giving a calculated three-photon Rabi frequency of $\sim90\,$kHz, in agreement with the measured value. 
The single-photon Rabi frequencies are measured via on-resonance excitation of the transition. 
The values of $\Delta$ and $\delta$ are large compared to the Doppler width of the atomic ensemble ($\sim60$\,kHz) to mitigate the effect of non-uniform three-photon Rabi frequency and AC Stark shift across different velocity classes~\cite{Grynberg1976}.
Applying a three-photon excitation pulse on resonance for various durations results in Rabi oscillations (Fig.~\ref{fig:fig689}d).
The offset in the $^3P_1$ population from three-photon excitation in Figs.~\ref{fig:fig689}c and \ref{fig:fig689}d results from off-resonance single-photon scattering which could be further suppressed by increasing the detunings $\Delta$ and $\delta$.

To characterize the Doppler sensitivity, we perform Ramsey spectroscopy using the pulse sequence in Fig.~\ref{fig:fig689ramsey}a. For various settings of the Ramsey delay duration $\tau$ we scan the relative phase $\phi$ between the Ramsey $\pi/2$ pulses (Fig.~\ref{fig:fig689ramsey}b). From a fit of the function $A+\frac{C}{2}\sin(\phi+\phi_{0})$ to the data, we can extract the Ramsey contrast $C$, with $A$ and $\phi_0$ as free parameters. For single-photon excitation in a 3\,$\mu$K cloud, the Ramsey contrast rapidly decays due to Doppler dephasing and is indistinguishable from zero beyond $\tau \sim 15$\,$\mu$s, while the three-photon Ramsey coherence $1/e$ decay time is 30(4)\,$\mu$s (Fig.~\ref{fig:fig689ramsey}c).  The temperature of the atomic ensembles can be increased up to 40$\,\mu$K by applying a 20-50\,$\mu$s, near-resonant 461\,nm pulses to heat the atoms with negligible change to the atom cloud size \cite{Anderegg2018}. Increasing the atomic cloud temperature results in an increased rate of single-photon Ramsey contrast decay. When the dominant decaying mechanism is Doppler dephasing rather than spontaneous emission from the excited state, the contrast scales as  $\ln C \propto -T\tau^2$, where $T$ is the temperature of the atomic ensemble and $\tau$ is the Ramsey delay duration, as shown in Ref.\,\cite{biedermann2012}. For experimentally optimized three-photon laser alignment, an increase in the cloud temperature from 3 to 40 $\mu$K  has only a slight effect on the decoherence. In this case, the contrast decay scales mostly exponentially with respect to $\tau$, consistent with coherence loss resulting primarily from spontaneous emission and other homogenous decay mechanisms. Deliberate misalignment of the three-photon lasers results in increased Doppler sensitivity (Fig.~\ref{fig:fig689ramsey}c). This misalignment introduces residual $k$-vector parallel to the atomic velocity that results in a residual Doppler shift associated with the three-photon process of $\Delta \vec{k} \cdot \vec{v} =  (\vec{k}_{1}-\vec{k}_{2}+\vec{k}_{3})\cdot \vec{v}$. This makes Doppler dephasing an increasingly significant decay mechanism, with a Doppler width that is $0.09$ and $0.13$ times that of the single-photon process for the cases of $\Delta\theta_{2}=4^{\circ}$ and $\Delta\theta_{1}=2.5^{\circ}$, $\Delta\theta_{2}=4^{\circ}$, respectively. Complete analysis of Ramsey experiment contrast decay is provided in the Appendix B.

\section{AC Stark shifts}
For a given set of detunings $\Delta, \delta$ and single-photon Rabi frequencies $\Omega_1$, $\Omega_2$, and $\Omega_3$, as defined in Fig.\,1a above, the total AC Stark shift $\delta_{AC}$ for the three-photon excitation can be approximated by the sum of the individual AC Stark shifts from each laser~\cite{Foot}
\begin{equation}\label{eq:AC}
    \delta_{AC} \approx \frac{\Omega_1^2}{2\Delta} +  \frac{\Omega_2^2}{2(\Delta-\delta)} -  \frac{\Omega_3^2}{2\delta}\, .
\end{equation}
Therefore, it is generally possible to find a set of single-photon Rabi frequencies that null the the total AC Stark shift without the use of additional fields that are not essential to the excitation process.

\begin{figure}[t!]
    \centering
    \includegraphics[width = 2.5in]{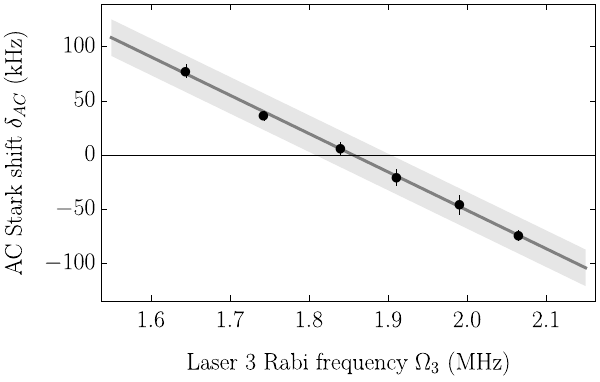}
    \caption{The AC Stark shift of the three-photon $^{1}S_{0}$\,$\leftrightarrow$\,$^{3}P_{1}$ excitation scheme versus laser 3 Rabi frequency with $\Omega_1 = 0.57(1)\,$MHz, $\Omega_2 =  2.04(1)\,$MHz, and detunings of $\Delta = 12\,$MHz and $\delta = 4\,$MHz. Data points are differences between single- and three-photon resonances offset by $\Delta$. Error bars indicate 95\% confidence intervals for the center of the three-photon resonance. The shaded area represents the standard error of the estimated single-photon resonance frequency. The solid line is a linear fit to the data.}
    \label{fig:ac}
\end{figure}

Operating with parameters of $\Omega_1 = 0.57(1)\,$MHz, $\Omega_2 =  2.04(1)\,$MHz, and detunings of $\Delta = 12\,$MHz and $\delta = 4\,$MHz, we are able to tune $\delta_{AC}$ to zero at $\Omega_3 \sim  1.85\,$MHz, as shown in Fig.~\ref{fig:ac}. 
The AC Stark shift is obtained by first measuring the single-photon resonance frequency with single-photon spectroscopy using laser 1 and all other lasers off. 
Then, the three-photon resonance frequency is measured by scanning the frequency of laser 1 with all three-photon lasers on using the parameters listed above. 
A Lorentzian function is fit to both resonance features to estimate their resonance frequencies.
The relative frequency difference between the single- and three-photon resonances minus $\Delta$ gives the AC Stark shift. The recoil shift is on the order of $\sim10$\,kHz and is within the standard error of our single-photon resonance estimate. 

The sensitivity of the AC Stark shift to laser intensity is a consideration for time keeping applications of these schemes. In general, the engineering control required to mitigate systematic offsets due to AC Stark shifts depends on the particular application.  Here we note that methods developed for precision atom interferometry and timekeeping are applicable to this system.  
From Fig.~\ref{fig:ac} below, we measure a fractional frequency shift $\delta_{AC}/\omega_0$ as a function of the power of laser 3 with the powers of laser 1 and 2 fixed. 
This results in a shift coefficient of $7 \times 10^{-11}$/mW.
The use of a single laser source for all three-photon lasers leads to largely common-mode intensity fluctuations in each laser path. 
Thus, when the three-photon laser detunings and Rabi frequencies are configured to null the AC Stark shift, fluctuations in the source laser's intensity result in a total AC Stark shift that remains nulled. 
Independent intensity fluctuations in the individual beam paths after splitting from the laser source lead to residual AC Stark shift sensitivity to laser intensity.
Such beam path fluctuations could result from residual etalon effects, polarization drift, and misalignment due to thermal effects.

In this demonstration we operate in a highly coherent regime with an excitation fraction on the order of unity. 
This can be relaxed and the lasers can be further detuned to decrease sensitivity to non-common intensity fluctuations. 
For instance, increasing the detunings used in Fig.~\ref{fig:ac} each by a factor of 30 would reduce $\delta_{\text{AC}}/\omega_0$ by a factor of 30 while maintaining a $\sim 1\%$ excitation fraction in the atomic sample.
Furthermore, ``auto-balancing'' techniques are available to mitigate AC stark effects in Ramsey spectroscopy (see, e.g. \cite{yudin2020}). These are applicable to three-photon schemes. Finally, the incorporation of additional non-resonant spectral components provides another route to mitigating AC Stark shifts.  These methods are commonly employed in precision atom interferometric sensors \cite{Asenbaum2020}.

\section{Excitation of the $^{3}P_{0}$ transition}
\begin{figure*}
\includegraphics[width=6.5in]{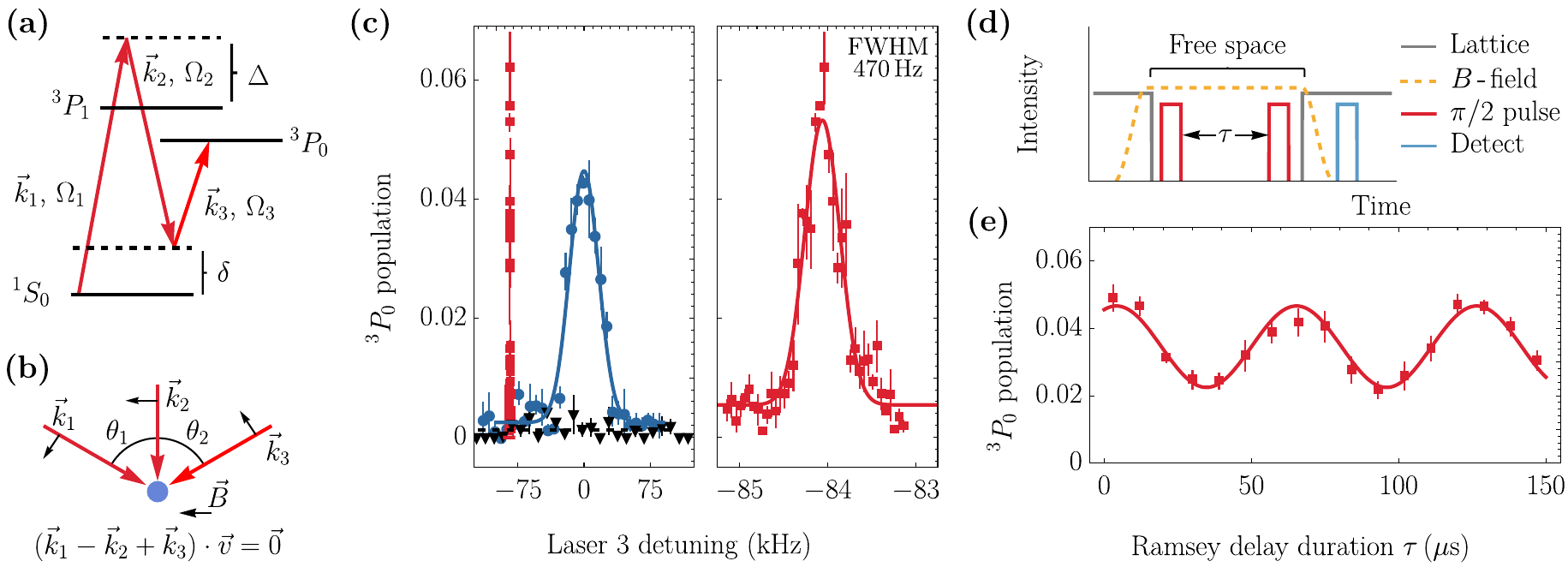}
\caption{\label{fig:fig698} Three-photon spectroscopy on the $^{1}S_{0}$\,$\leftrightarrow$\,$^{3}P_{0}$~clock transition. a)  Energy level diagram (not to scale) for three-photon excitation of $^{1}S_{0}$\,$\leftrightarrow$\,$^{3}P_{0}$. b)  Laser geometry for Doppler-free operation. The three lasers have wavevectors $\vec{k}_{i}$, $i\in \{1,2,3\}$ and propagate in a plane, with lasers~1 and 2 separated by angle $\theta_{1}$ and lasers~2 and 3 separated by $\theta_{2}$. The laser polarization vectors (indicated by black arrows) and the magnetic field $\vec{B}$ all lie in the plane containing the laser propagation vectors. c) Measured $^{3}P_{0}$ population after $4$\,ms of free-space interrogation as the frequency of laser~3 is varied relative to the atomic resonance (0\,kHz).
Single-photon excitation at Rabi frequency $\sim$700\,Hz (blue circles) results in a Doppler-broadened profile with full-width-half-maximum of 45\,kHz, while single-photon excitation with Rabi frequency $\sim$70\,Hz (black triangles) results in no significant population transfer. Three-photon Doppler-free excitation with Rabi frequency $\sim$70\,Hz (red squares) reduces the spectroscopic linewidth 100--fold. Data points and error bars are averages and standard errors of 3 repeated measurements. Solid curves are Gaussian fits to the data.
d) Experimental sequence for free-space Ramsey spectroscopy. Three-photon Ramsey pulses of duration $2$\,ms are separated by a variable delay duration $\tau$. See text for further details. 
e) Ramsey oscillations for three-photon excitation of the $^{1}S_{0}$\,$\leftrightarrow$\,$^{3}P_{0}$ transition as the delay duration $\tau$ is varied. Data points and error bars are averages and standard errors of 9 repeated measurements. The solid curve is a sinusoidal fit to the data. }
\end{figure*}

Three-photon excitation of the $^{1}S_{0}$\,$\leftrightarrow$\,$^{3}P_{0}$ transition is accomplished using the scheme shown in Figs.~\ref{fig:fig698}a and \ref{fig:fig698}b. Lasers~1 and 2, near resonance with the $^{1}S_{0}$$\leftrightarrow\, $$^{3}P_{1}$ transition, are derived from a single 689 nm source that is split into two paths. This configuration enables suppression of relative frequency fluctuations between the lasers. Laser 3, near resonance with the 698\,nm  $^{1}S_{0}$\,$\leftrightarrow$\,$^{3}P_{0}$ transition, is frequency-offset locked to the 689\,nm source via an optical transfer cavity. A 375\,G static magnetic field applied parallel to the polarization of laser 2 enables optical coupling between the $^{1}S_{0}$ and $^{3}P_{0}$ states \cite{Taichenachev2006}. The polarization vectors of all lasers lie in the plane containing the laser propagation vectors. As before, AOMs in each laser path allow for precise control of the amplitude, frequency and relative phase between each of the lasers. At the location of the atoms, lasers 1, 2, and 3 are collimated with waists of 0.8\,mm, 0.8\,mm, and 0.5\,mm respectively. 

We initially prepare a $\sim3\,\mu$K  cloud of atoms as described above, then transfer the atoms into a $\sim15\,\mu$K-deep one-dimensional 813\,nm optical lattice.
The lattice is employed to prevent the expansion of the atoms for the 30\,ms required to switch on the coupling magnetic field, and reduces the transverse spatial extent of the cloud by a factor of $\sim\,2$ with negligible change in the atomic temperature.  
The lattice laser is perpendicular to the plane of the three-photon lasers (Fig~\ref{fig:fig698}b).
The lattice is then switched off during the three-photon excitation and the atoms undergo free-space expansion. All three-photon lasers are applied simultaneously to excite the atoms. Figure~\ref{fig:fig698}c shows the three-photon resonance spectrum obtained with $\Omega_1 = 1.67(5)\,$MHz, $\Omega_2 = 2.80(5)\,$MHz, and an inferred $\Omega_3 = 700\,$Hz given the size of laser 3 and coupling magnetic field\,\cite{Taichenachev2006}. We select $\Delta = 185\,$MHz to minimize single-photon scattering and $\delta=70\,$kHz to shift the three-photon resonance beyond the single-photon $^{1}S_{0}$\,$\leftrightarrow$\,$^{3}P_{0}$ Doppler width. Note that if $\delta$ were within the Doppler width of the atomic sample, atoms of different velocity classes would experience a non-negligible variation in three-photon Rabi frequency resulting in significant inhomogeneous broadening, as discussed in Ref.~\cite{Grynberg1976}.  Moreover, we choose $\Delta$ and $\delta$ such that $\Omega_1^2/2\Delta$, $\Omega_2^2/2\Delta$, and $\Omega_1\Omega_2/2\Delta$ are all smaller than $\delta$.
The parameters selected here result in a three-photon Rabi frequency of $\Omega_{3v}\sim 70\,$ Hz, a factor of 10 lower than the single-photon Rabi frequency of the $^{1}S_{0}$\,$\leftrightarrow$\,$^{3}P_{0}$ transition at the same laser intensity and magnetic field. 

After free-space interrogation, the atoms are recaptured in the lattice while the mixing magnetic field is ramped down before state readout. Approximately $50$\% of the atoms are recaptured, resulting in a single-photon Doppler width of 45\,kHz for the detected ensemble, as shown in Fig.~\ref{fig:fig698}b. 
Experimental optimization of the laser angles results in a two orders of magnitude reduction in the spectroscoptic linewidth of the atom cloud (Fig.~\ref{fig:fig698}), down to 470\,Hz. For a direct comparison of the Doppler sensitivity between single-photon and three-photon excitation, we perform single-photon interrogation with the single-photon Rabi frequency reduced to that of the three-photon process by decreasing the mixing magnetic field by a factor of 10.
In this case, we see negligible population of the $^3P_0$ state, while the three-photon process at the same Rabi frequency populates a substantial fraction of the atom cloud, as shown in Fig.~\ref{fig:fig698}c.  We perform free-space Ramsey spectroscopy with three-photon excitation on this transition and observe Ramsey oscillations (Fig.~\ref{fig:fig698}e). 
The decrease in excitation fraction compared to the $^{1}S_{0}$\,$\leftrightarrow$\,$^{3}P_{1}$ scheme is due to the atoms freely falling out of the laser interaction region during the three-photon interrogation sequence. 

\section{Clock applications}
Here we project possible performance for atomic clocks based on three-photon excitation with narrow-line transitions. Stability projections assume that the system's noise is limited by atomic projection noise for a given atomic flux. 
We also assume that the three-photon lasers are sufficiently well aligned to ensure that stability is not limited by residual first-order Doppler shifts.
Repeatability and accuracy projections are based on assumptions associated with specific atomic clock configurations, as detailed below.  Specifically, we consider errors induced by AC Stark shifts resulting from uncontrolled variation in the laser intensities used to drive the three-photon transitions.  We also estimate inaccuracy due to residual first-order Doppler shifts resulting from imperfect alignment of the lasers.  In the analysis below, our intent is not to provide an exhaustive analysis; rather, it is to illustrate the potential associated with three-photon Doppler-free excitation and to serve as a catalyst for future work. 

Two configurations are considered:  1) a thermal atomic beam operated on the $^{1}S_{0}$\,$\leftrightarrow$\,$^{3}P_{1}$ transition and 2) a laser-cooled fountain operated on the $^{1}S_{0}$\,$\leftrightarrow$\,$^{3}P_{0}$ transition.

\begin{enumerate}
\item {\it Thermal atomic beam.}  The fractional frequency instability is estimated from the transition $Q$ and the atomic projection noise associated with the detected atom flux.  Assuming a detected flux of $10^{10}$ atoms/s, the projected instability is $\sim 2\times10^{-16}/\sqrt{\text{Hz}}$ for the $^{1}S_{0}$\,$\leftrightarrow$\,$^{3}P_{1}$ transition.  We assume repeatability errors from the first-order Doppler effect to be given by unknown errors in laser alignment.  As a rough model, we assume the lasers are delivered from a thermally stable optics bench which is temperature controlled to achieve $< \,$ 1 nrad uncompensated angle alignment variations.  For an atomic beam operating with a most probable velocity of $\sim$ 360 m/s (corresponding to temperature sufficient to meet the detected flux above), the resulting repeatability error is $\sim 1\times10^{-15}$.  The AC Stark shift-induced fractional frequency instability is estimated to be $< \, 2\times10^{-16}$ assuming intensities $I$ are controlled at the $\delta I/I < 1 \times 10^{-4}$ level and a factor of 100$\times$ further suppression is provided by the AC Stark nulling and intensity modulation methods described above.  This estimate follows from the data presented in Fig. 3.

\item {\it Cold atomic fountain}.  Cold Sr atomic ensembles of $N = 10^{6}$ atoms with $T_c$ = 10 ms interrogation times are sampled in an interleaved, zero-dead-time configuration \cite{Biedermann2013}.  The stability is $\sim \frac{1}{Q}\sqrt{\frac{T_c}{N\tau}}\,$, where $\tau$ is the averaging time~\cite{Enzer21}, giving $\sim 1\times 10^{-17}/\sqrt{\text{Hz}}$.  Accuracy errors due to first-order Doppler shifts result from both the atom ensemble mean velocity drifts and the angle errors in laser alignment.  Assuming a mean velocity drift of $< 100 \, \mu$m/s and $< 100 \, \mu$rad laser beam alignment errors, the estimated fractional frequency error from this source is $\sim 3 \times 10^{-17}$. The first-order Doppler errors associated with the acceleration due to gravity are largely suppressed by the fountain geometry.  The limit on accuracy from the AC Stark effect is estimated to be $< 2\times10^{-17}$, again assuming intensities $I$ are controlled at the $\delta I/I < 1 \times 10^{-4}$ level and a factor of 100$\times$ further suppression from the methods described above.  The AC Stark shift sensitivity is roughly a factor of 10 reduced from that of Fig. 3 above for the $^{1}S_{0}$\,$\leftrightarrow$\,$^{3}P_{0}$ transition due to the lower required Rabi frequency.

\end{enumerate}

Both configurations above compare favorably to many modern atomic frequency standards~\cite{Marlow20}.

\section{Conclusion}
In conclusion, we have demonstrated Doppler-free and recoil-free three-photon excitation of narrow-line transitions in atomic strontium. This method enables spectroscopy of odd-parity transitions and participation of all atoms in a thermal ensemble, potentially allowing more atomic or molecular species to be used for high-precision spectroscopy in situations where confinement or cooling is challenging or undesirable. In particular, our techniques could be applied directly to other narrow-line transitions found in alkaline-earth-like atoms used in many of the most precise atomic clocks. Three-photon excitation schemes have the potential to facilitate the development of robust and compact thermal-beam atomic frequency standards providing an alternative to Ramsey-B\'orde based systems that use narrow-line transitions in calcium~\cite{olson2019ramsey}. In comparison to two-photon $^{87}$Rb optical clocks using the $^2S_{1/2}$\,$\leftrightarrow$\,$^2D_{5/2}$ transition, which are limited by a $\sim 300$\,kHz linewidth, a three-photon Doppler-free approach using the $^{1}S_{0}$\,$\leftrightarrow$\,$^{3}P_{1}$ transition in atomic strontium or calcium would benefit from  the $7.5$\,kHz and $470$\,Hz natural linewidths, respectively. 
These methods could provide an alternative to current narrow-line and clock transition excitation schemes while maintaining wide scientific and technological reach.

\begin{acknowledgments}
We wish to thank Jason Hogan, Jan Rudolph, and Leo Hollberg for helpful conversations. G.P. acknowledges support from the U.S. National Science Foundation GRFP.
\end{acknowledgments}

\appendix

\section{Three-photon Rabi frequency}
The three-photon Rabi frequency for the $^{1}S_{0}$\,$\leftrightarrow$\,$^{3}P_{1}$ excitation scheme is a factor of 2 larger than the expression for the three-photon Rabi frequency in a ladder system~\cite{Ryabtsev2011}.  This is because there are two resonant three-photon excitation paths for the $^{1}S_{0}$\,$\leftrightarrow$\,$^{3}P_{1}$ system, shown in Fig.~\ref{fig:paths} below. 
The Rabi couplings from these two paths add.  The resulting analytic expression for the Rabi frequency agrees with numerical simulations of the system.
\begin{figure}[h!]
    \centering
    \includegraphics[width = 3 in]{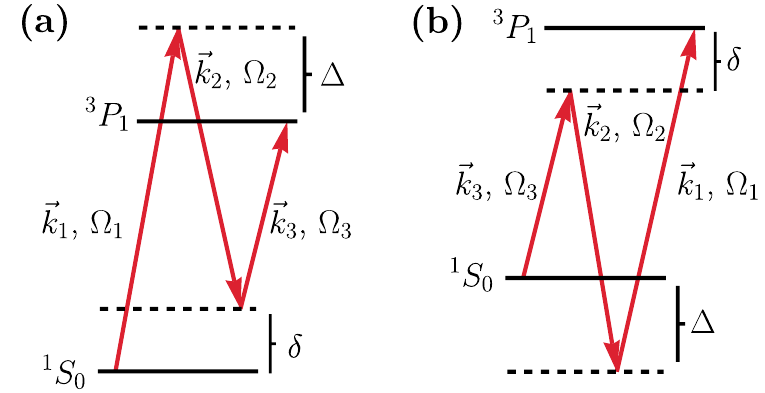}
    \caption{Three-photon excitation paths for the  $^{1}S_{0}$\,$\leftrightarrow$\,$^{3}P_{1}$ transition. The wavevector and Rabi frequency for laser $i\in\{1,2,3\}$ is given by $\vec{k}_i$ and $\Omega_i$, respectively. Here $\Delta$ and $\delta$ indicate detunings from the $^{1}S_{0}$\,$\leftrightarrow$\,$^{3}P_{1}$ resonance frequency.}
    \label{fig:paths}
\end{figure}

\section{Characterization of Doppler sensitivity}  
Here we describe the fitting procedure used to quantify Ramsey contrast decay in the three-photon $^{1}S_{0}$\,$\leftrightarrow$\,$^{3}P_{1}$ excitation shown in Fig.~\ref{fig:fig689ramsey}c.

We parameterize the Ramsey contrast $C$ as a function of the Ramsey delay time $\tau$ using the expression
\begin{equation}
    \ln C (\tau) \propto -\left(\frac{\tau}{t_D}\right)^2 - \frac{\tau}{t_1} + y_0,
\end{equation}

\noindent
where the first term on the right is a Gaussian decay process with characteristic decay time $t_D$ due to Doppler dephasing. The second term is an exponential decay process with characteristic decay time $t_1$. The constant offset $y_0$ relates to the $\pi$-pulse fidelity.
For a single-photon process, the dominant decay mechanism is Doppler dephasing due to the temperature of the atom ensemble. 
For an optimized three-photon process, the Gaussian decay contribution becomes negligible and the dominant decay mechanisms are spontaneous emission from the excited state and other homogeneous decay processes resulting in predominantly exponential contrast decay.

We fit this decay function to the contrast decay data to extract the characteristic decay times. The characteristic dephasing time $t_D$ is inversely proportional to the Doppler width $\Gamma_D$ of the ensemble~\cite{biedermann2012}, thereby scaling with temperature and laser $k$-vector as
\begin{equation}
    t_D \propto \frac{1}{\Gamma_D} \propto \frac{1}{|\Delta \vec{k}|\sqrt{T}}
\end{equation}
where $T$ is the temperature of the atom cloud and $|\Delta \vec{k}|$ is the magnitude of the residual $k$-vector for the process. For three-photon excitation, $\Delta \vec{k} = \vec{k}_1-\vec{k}_2+\vec{k}_3$ while for the single-photon process $\Delta \vec{k} = \vec{k}$, where $\vec{k}$ is the $k$-vector resonant with the $^{1}S_{0}$\,$\leftrightarrow$\,$^{3}P_{1}$ transition, of nearly equal magnitude to each $\vec{k}_i$ in the three-photon excitation. 
It follows that for known temperature $T$, the time constant $t_D$ relates directly to $\Delta \vec{k}$. 
Comparing $t_D$ between single-photon and three-photon excitation configurations allows us to quantify the level of Doppler sensitivity, which we define as $|\Delta\vec{k}|/|\vec{k}|$. 

For each excitation configuration shown in Fig.~2c -- single-photon, three-photon optimized, three-photon $\Delta\theta_1 = 4^{\circ}$, and three-photon $\Delta\theta_1 = 4^{\circ}, \Delta\theta_2 = 2.5^{\circ}$ -- we perform a combined least-squares fit to the log-contrast data across all temperatures $T_i \in \{3\,\mu\text{K},13\,\mu\text{K},40\,\mu\text{K}\}$ using the fit function 
\begin{equation}
    f_i(\tau; \alpha, \beta, y_{0,T_i}) = \alpha T_i \tau^2 + \beta \tau +y_{0,T_i},
\end{equation}
where free parameters $\alpha,\,\beta$ are optimized across all temperatures and the free parameter $y_{0,T_i}$ is optimized for a specific atom cloud temperature $T_i$. 
We denote the optimal fit parameters by $\textbf{v}^* = \{\alpha^*, \beta^*, y_{0,3\mu K}^*, y_{0,13\mu K}^*, y_{0,40\mu K}^*\}$ and use their values to estimate the underlying Doppler dephasing and exponential decay times across all temperatures of a particular configuration. The $95\%$ confidence intervals for all optimal parameters of our fits in all excitation configurations are provided in Table~\ref{tab:table1}. 
Note that for all three-photon excitation configurations, the values of $\beta^*$ are not significantly distinguishable from each other, suggesting the only major disparity between their decay processes is the variation in $|\Delta \vec{k}|$ from deliberate misalignment of the three-photon lasers.

\begin{figure*}
    \includegraphics[width = 5 in]{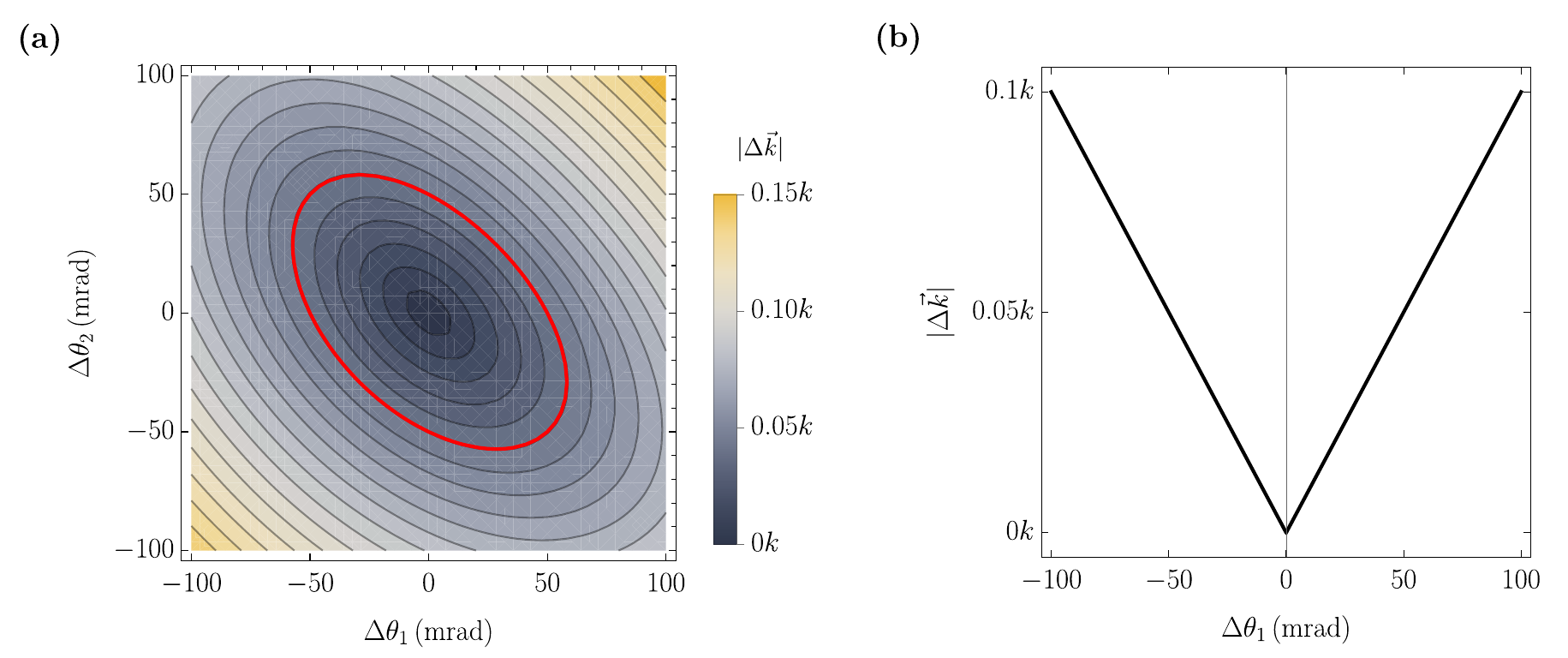}
    \caption{Contour plot of the magnitude of the residual $k$-vector $|\Delta \vec{k}|$ of the three-photon excitation as a function of the angle deviations from the nulled orientation for the $^{1}S_{0}$\,$\leftrightarrow$\,$^{3}P_{1}$ excitation scheme. (a) Contour plot for residual $k$-vector magnitudes. Red contour is the measured residual $k$-vector magnitude ($|\Delta \vec{k}| = 0.05$) for the experimentally-optimized configuration.
    (b) Plot of the residual $k$-vector magnitude of the three-photon process as a function of the angle deviation $\Delta \theta_1$ from the ideal Doppler-free condition with $\Delta \theta_2$ fixed at zero. }
    \label{fig:ang_dev}
\end{figure*}

\begin{table*}
\footnotesize
\caption{\label{tab:table1}
Table of 95\% confidence intervals for optimal fit parameters $\textbf{v}^*$ for Ramsey experiment contrast decay from single- and three-photon (TP) excitation configurations}\begin{ruledtabular}
\begin{tabular}{cccccc}
 Excitation configuration & $\alpha^* \times 10^{5}$ & $\beta^* \times 10^{2}$ & $y_{0,3\mu K}^*$ & $y_{0,13\mu K}^*$ & $y_{0,40\mu K}^*$ \\
\hline
TP optimized & $(-1.98, -0.21) $ & $(-3.75, -2.88)$ & $(-1.09, -0.98)$ & $(-1.11, -1.01)$ & $(-1.17, -1.03)$ \\
TP $\Delta\theta_2 = 4^{\circ}$ & $(-4.74, -3.36) $ & $(-3.86, -3.08)$ & $(-1.09, -1.00)$ & $(-1.09, -1.02)$ & $(-1.18, -1.11)$ \\
TP $\Delta\theta_2 = 4^{\circ}$, $\Delta\theta_1 = 2.5^{\circ}$ & $(-9.35, -6.41)$ & $(-4.61, -3.32)$ & $(-1.14, -0.97)$ & $(-1.10, -1.03)$ & $(-1.19, -1.09)$  \\
Single-photon & $(-514, -438)$ & $(-4.84, 0.92)$ & $(-0.75, -0.54)$ & $(-0.77, -0.64)$ & $(-0.85, -0.72)$ \\
\end{tabular}
\end{ruledtabular}

\end{table*}

By comparing the value of $\alpha^*$ obtained by the fitting procedure for each excitation scenario to that of the single-photon case ($\alpha^*_{1\nu}$), we compute $|\Delta \vec{k}|/|\vec{k}|$ to quantify the Doppler sensitivity of each alignment configuration, shown in Table~\ref{tab:table2}.
These estimates are comparable to the values of the residual $k$-vector calculated from the angle deviations $\{\Delta\theta_i\}$ deliberately introduced assuming perfect Doppler-free alignment in the experimentally optimized three-photon configuration.
This analysis also provides an estimate of $|\Delta \vec{k}|/|\vec{k}|$ for the  experimentally-optimized three-photon configuration,
where residual $k$-vector is still present at low levels, with an estimated $|\Delta \vec{k}|/|\vec{k}|$ of 0.05(3). This results in a Doppler dephasing time $t_D$ that only becomes comparable to the decay time from homogeneous mechanisms $t_1$ at atom cloud temperatures~$>40\,\mu$K. 

\begin{table*}
\caption{\label{tab:table2}
Estimates for residual Doppler sensitivity $|\Delta\vec{k}|/|\vec{k}|$. Estimates are obtained from optimized fit parameter $\alpha^*$ (middle column) and from angle deviations introduced (right column) for each three-photon excitation configuration explored above. 
}\begin{ruledtabular}
\begin{tabular}{ccc}
 Three-photon configuration & Estimated $|\Delta\vec{k}|/|\vec{k}|$ via $\sqrt{\alpha^*/\alpha^*_{1\nu}}$ & Estimated $|\Delta\vec{k}|/|\vec{k}|$ via $\{\Delta\theta_i\}$ \\
\hline
Optimized & $0.05\pm0.03$ & 0\footnote{Experimentally-optimized configuration and reference point for angle misalignment.} \\
$\Delta\theta_2 = 4^{\circ}$ & $0.09\pm0.01$ & $0.07\pm0.01$  \\
$\Delta\theta_2 = 4^{\circ}, \Delta\theta_1 = 2.5^{\circ}$ & $0.13\pm0.01$& $0.10\pm0.02$ \\
\end{tabular}
\end{ruledtabular}
\end{table*}

From the fit function evaluated using the optimized fit parameters $\textbf{v}^*$, we calculate 95\% confidence intervals for the $1/e$ contrast decay time, shown in Table~\ref{tab:table3}.
Note that for the optimally aligned three-photon process, these confidence intervals overlap, suggesting insignificant change in contrast decay time for a temperature range that varies by over an order of magnitude. 
This is not true for any of the other excitation configurations explored.

\begin{table*}
\caption{\label{tab:table3}
Table of 95\% confidence intervals for the Ramsey experiment $1/e$ decay time in $\mu$s of each single- and three-photon (TP) excitation configuration and atom cloud temperature explored above.}\begin{ruledtabular}
\begin{tabular}{cccc}

 Configuration / Temperature & $3\mu$K & $13\mu$K & $40\mu$K \\
\hline
TP optimized &  (22.32, 41.32) & (25.78, 29.54)  & (20.45, 25.83)  \\
TP $\Delta\theta_2 = 4^{\circ}$ & (24.24, 30.90) & (18.21, 23.10)  & (15.79, 16.93) \\
TP $\Delta\theta_2 = 4^{\circ}$, $\Delta\theta_1 = 2.5^{\circ}$ & (21.73, 27.68) & (15.93, 18.07)  & (12.40, 13.05) \\
Single-photon & (6.45, 6.86) & (3.49, 3.65)  & (2.07, 2.17) \\
\end{tabular}
\end{ruledtabular}
\end{table*}

\begin{table*}
\caption{\label{tab:table4}
Table of alignment accuracy required for three-photon lasers for operation at various atomic ensemble temperatures. The residual three-photon Doppler width for $1\,$arcsec alignment accuracy is provided for each atomic ensemble temperature considered. }\begin{ruledtabular}
\begin{tabular}{cccc}

Temperature & 10\,$\mu$K & 1\,mK  & 1\,K \\ 
\hline
Alignment accuracy for 7.6\,kHz Doppler width & 72\,mrad & 7.2\,mrad & 0.23\,mrad  \\
Alignment accuracy for 10\,Hz Doppler width & 95.4\,$\mu$rad & 9.5\,$\mu$rad  & 0.3\,$\mu$rad \\
Doppler width at $1\,$arcsec alignment accuracy & 0.5\,Hz & 5.2\,Hz & 166\,Hz \\
\end{tabular}
\end{ruledtabular}
\end{table*}

The dependence of the magnitude of the residual $k$-vector $|\Delta \vec{k}|$ on laser angle deviation from the Doppler-free orientation is shown in Fig.~\ref{fig:ang_dev}a.
Observe that moderately-aligned lasers ($\Delta\theta_i\sim$1\,mrad) will reduce the Doppler-broadened spectroscopic linewidth by a factor of $\sim$1000. Achieving $<1\,$arcsec accuracy for laser alignment is fairly routine~\cite{olson2019ramsey}.
Near the Doppler-free configuration, the sensitivity of $|\Delta \vec{k}|$ to fluctuations in the laser angles is given by $d |\Delta \vec{k} |/ d\Delta\theta_i=1\,k/$rad, as shown in Fig.~\ref{fig:ang_dev}b, and the residual three-photon Doppler width is $\Gamma_D \times \Delta\theta$.

For hotter atomic samples in, for example, MOTs using dipole-allowed transitions and thermal atomic beams, alignment accuracy becomes an increasingly stringent requirement for operation.
In Table~\ref{tab:table4}, we show the laser alignment accuracy required for operation in atomic samples of temperatures $10\,\mu$K (narrow-line MOTs), $1\,$mK (dipole-allowed MOTs), and $1\,$K (thermal atomic beams) as well as the residual three-photon Doppler width for $1\,$arcsec alignment accuracy. 


\bibliography{ref}

\end{document}